\documentclass{article}[12pt]
\usepackage{amsmath,amssymb,times,epsfig,graphics} 
\begin{document}
\vspace{.5in}

\def\tb{\widetilde{b}}
\def\tf{\widetilde{f}}
\def\tp{\widetilde{p}}
\def\tq{q}

\title{Analytical Mechanics in Stochastic Dynamics:
Most Probable Path, Large-Deviation Rate Function 
and Hamilton-Jacobi Equation}

\author{Hao Ge$^1$ and Hong Qian$^{2}$\\
\\[0.3cm]
$^1$Beijing International Center for Mathematical Research 
(BICMR)\\
and Biodynamic Optical Imaging Center (BIOPIC)
\\ 
Peking University, Beijing, 100871 PRC\\[0.3cm]
$^2$Department of Applied Mathematics\\
University of Washington, Seattle, WA 98195, USA
}

\maketitle

\tableofcontents

\begin{abstract}
Analytical (rational) mechanics is the mathematical structure of
Newtonian deterministic dynamics developed by  D'Alembert,
Langrange, Hamilton, Jacobi, and many other luminaries 
of applied mathematics.  Diffusion as a stochastic process of
an overdamped individual particle immersed in a fluid, 
initiated by Einstein, Smoluchowski, Langevin and Wiener,
has no momentum since its path is nowhere differentiable. 
In this exposition, we illustrate how 
analytical mechanics arises in stochastic dynamics from a 
randomly perturbed ordinary differential equation $dX_t=b(X_t)dt+\epsilon dW_t$ where $W_t$ is a Brownian
motion.  In the limit of vanishingly small $\epsilon$, 
the solution to the stochastic 
differential equation other than $\dot{x}=b(x)$ are all
rare events.  However, conditioned on an occurence of such
an event, the most probable trajectory of the stochastic
motion is the solution to Lagrangian mechanics with
$\mathcal{L}=\|\dot{q}-b(q)\|^2/4$ and
Hamiltonian equations with $H(p,q)=\|p\|^2+b(q)\cdot p$. 
Hamiltonian conservation law implies
that the most probable trajectory for a ``rare'' event has 
a uniform ``excess kinetic energy'' along its path.  Rare
events can also be characterized by the principle of large
deviations which expresses the probability density function for 
$X_t$ as $f(x,t)=e^{-u(x,t)/\epsilon}$, where
$u(x,t)$ is called a large-deviation rate function which
satisfies the corresponding Hamilton-Jacobi equation.
An irreversible diffusion process with $\nabla\times b\neq 0$
corresponds to a Newtonian system with a Lorentz force
$\ddot{q}=(\nabla\times b)\times \dot{q}+\frac{1}{2}\nabla\|b\|^2$.
The connection between stochastic motion and analytical 
mechanics can be explored in terms of various techniques 
of applied mathematics, for example, singular perturbations, viscosity solutions, and integrable systems. 
\end{abstract}

\section{Introduction}

{\em Dynamics} as an analytical concept is one of the
most important contributions of mathematics to modern
thinking. Currently there are three fundamentally 
different types of dynamics: classical, quantum,
and stochastic \cite{ao_ctp}. Quantum dynamics
aside, classical dynamics in term 
of deterministic ``trajectories'' of a system, 
continuous or not, is still the dominant
model in quantitative science and engineering.
However, rapid development of quantitative biology
in recent years, and the sustained interests in
statistical physics, has pushed the third, stochastic
dynamics, to the forefront of applied mathematics.
For recent reviews on {\em Darwinian dynamics} and
the {\em Delbr\"{u}ck-Gillespie process} for cellular
biochemical systems, see \cite{ao_ctp,qian_nonl}.
Novel and even non-orthodox stochastic dynamic
approaches to quantum phenomena can be found 
in \cite{chung,nagasawa,albeverio}.

    Describing the celestial mechanical system of
a few interacting bodies, Newton's equations of motion
in terms of classical dynamics is one of the most
successful mathematical models known to mankind.
An in-depth study of the subject in terms of 
{\em analytical mechanics} exposes
one to a wide range of applied mathematical theories
and techniques: Hamiltonian systems, Lagrangian
principle, and Hamilton-Jacobi equations are
several shining jewels of the treasure
box \cite{goldstein_book,landau_book,arnold_book}.

    In this exposition, we shall illustrate that much of these
classical, deterministic mathematics also emerge naturally 
in the theory of stochastic dynamics, when one is
interested in the relationship between stochastic
and deterministic dynamics.  This is reminiscent of
the semi-classical theory of quantum dynamics
developed in the 1970s \cite{miller}.  More importantly,
not only are they useful as tools for solving
problems, several quantities also acquired 
a strong probabilistic meaning, such as {\em large
deviation rate functions} and {\em most
probable paths}.

    Even though stochastic dynamics is often
described in terms of its probability distribution
changing with time, we need to emphasize, at the onset,
that neither the distribution perspective, nor a
trajectory perspective, is a complete description
of a stochastic process.  The mathematical
notion of a {\em random variable}, developed
by A. Kolmogorov, can not rest solely on
its distribution function, nor its
realizations \cite{kolmogorov}.  It is truly an
independent new mathematical object with deep
philosophical consequences.\footnote{Two issues
immediately come to mind:
($i$) The mathematical theory of probability requires
{\em all} possible outcomes being known {\em a
priori}, in the very definition of a random
variable.  This makes the concept of a
random variable only retrospective.  In statistics,
this is intimately related to the concept of
a {\em prior distribution}; and in economics this
distinguishes {\em risk} from {\em uncertainty}.
($ii$) Classical
dynamics has trajectories but only singular
distributions; quantum dynamics has distributions
but no trajectories due to Heisenberg's uncertainty
principle; stochastic dynamics requires
both perspectives.}

\section{Stochastic Dynamics in Terms of
Stochastic Differential Equations and Diffusion 
Processes} 

    One of the extensively studied problems
that connect stochastic and deterministic dynamics is
an ordinary differential equation with small
random perturbations \cite{freidlin_book}.   Let
us consider a diffusion process $X_t$ with
the stochastic differential equation
\begin{equation}
    dX_t = b(X_t)dt + \sqrt{2\epsilon}dB_t,
\label{1}
\end{equation}
and the corresponding Kolmogorov forward equation
\begin{equation}
    \frac{\partial f(x,t)}{\partial t}
     = \epsilon \frac{\partial^2 f}{\partial x^2}
     - \frac{\partial}{\partial x}\left( b(x) f(x,t)\right).
\label{2}
\end{equation}

In the limit of $\epsilon = 0$, 
(\ref{2}) is formally reduced to the first-order
partial differential equation (PDE):
\begin{equation}
    \frac{\partial f(x,t)}{\partial t}
     = - \frac{\partial}{\partial x}\left( b(x) f(x,t)\right),
\label{3}
\end{equation}
which is equivalent to, according to Liouville's theorem, 
a nonlinear ordinary differential
equation (ODE)
\begin{equation}
    \frac{dx(t)}{dt} = b\left(x(t)\right).
    \label{4}
\end{equation}
It is known that on any finite time interval $t\in[0,T]$,
the solution to Eq. (\ref{1}) with initial value
$X_0=x_o$ approaches the solution of the ODE with
probability 1 \cite{freidlin_book}.

    If we write $f(x,t)=e^{w(x,t)}$, then the linear
PDE (\ref{3}) becomes
\begin{equation}
    \frac{\partial w(x,t)}{\partial t}
     = - b(x)\frac{\partial w(x,t)}{\partial x}
    -\frac{db(x)}{dx},
\label{5}
\end{equation}
a linear, first-order partial differential equation.

Now for Eq. (\ref{2}), let us assume
$f(x,t)=e^{-u_{\epsilon}(x,t)/\epsilon}$, widely known as
the WKB ansatz.  Then we have
\begin{equation}
   \frac{\partial u_{\epsilon}(x,t)}{\partial t}
    = -\left[\left(\frac{\partial u_{\epsilon}}{\partial x}\right)^2
           + b(x)\frac{\partial u_{\epsilon}}{\partial x} \right]
      + \epsilon\left[\frac{\partial^2 u_{\epsilon}}{\partial x^2}
          + \frac{db(x)}{dx}\right].
\label{6}
\end{equation}
Therefore, to leading order, if $\lim_{\epsilon\rightarrow 0}
u_{\epsilon}(x,t)=u(x,t)$ exists and is differentiable, one has
\begin{equation}
   \frac{\partial u(x,t)}{\partial t}
    = -\left(\frac{\partial u(x,t)}{\partial x}\right)^2
           - b(x)\frac{\partial u(x,t)}{\partial x}.
\label{7}
\end{equation}
Eq. (\ref{7}) is widely called the Hamilton-Jacobi,
or Eikonal, equation (HJE) \cite{ross_jcp,saakian}. Note
that this equation is different from Eq. (\ref{5}): It is
a nonlinear, first-order PDE.  As we shall discuss
below, the solution to
(\ref{5}) is the limit of equation (\ref{2})
when $\epsilon\rightarrow 0$.  The solution
to (\ref{7}) is the {\em convergence rate} of that
limiting process. $u(x,t)$ is called the large-deviation
rate function.

	To relate a continuous random variable $X$ with 
probability density function $f_X(x)$ to a deterministic
quantity, the expected value $E[X]$ and modal value
$x^*$, with $f(x^*)\ge f(x)$, are often taken as the
counterpart.  Let us now consider a local minimum of 
the function $u(x,t)$ located at $x=x^*(t)$.  How does
the location and the value of the minimum change with time?
According to Eq. (\ref{7}):
\begin{eqnarray}
      \frac{du(x^*(t),t)}{dt} &=& \left[\frac{\partial u(x,t)}
         {\partial t} + \left(\frac{\partial u(x,t)}
         {\partial x}\right)
         \frac{dx^*(t)}{dt}
       \right]_{x=x^*(t)}
\label{8}\\[7pt]
    &=& \left[-\left(\frac{\partial u(x,t)}
           {\partial x}\right)^2
           - b(x)\frac{\partial u(x,t)}{\partial x}
       + \left(\frac{\partial u(x,t)}{\partial x}\right)
         \frac{dx^*(t)}{dt}
       \right]_{x=x^*(t)} 
\nonumber\\
     &=& 0.
\nonumber
\end{eqnarray}
Also from Eq. (\ref{7}) we have:
\begin{equation}
   0 = \frac{d}{dt}\left(
        \frac{\partial u(x^*(t),t)}{\partial x}
        \right)
    = \left[ \frac{\partial^2 u(x,t)}{\partial x\partial t}
         +\frac{\partial^2 u(x,t)}{\partial x^2}
         \left(\frac{dx^*(t)}{dt}\right) \right]_{x=x^*(t)}.
\end{equation}
Therefore,
\begin{eqnarray}
    \frac{dx^*(t)}{dt} &=& -\left[\left(
        \frac{\partial^2 u(x,t)}{\partial x^2}\right)^{-1}
         \frac{\partial^2 u(x,t)}{\partial x\partial t}
           \right]_{x=x^*(t)}
\nonumber\\[8pt]
    &=& \left[\left(
        \frac{\partial^2 u(x,t)}{\partial x^2}\right)^{-1}
        \left(2\frac{\partial u(x,t)}{\partial x}
               \frac{\partial^2 u(x,t)}{\partial x^2}+
                \frac{db(x)}{dx}\frac{\partial u(x,t)}{\partial x}
         \right.\right.
\nonumber\\
           && \left.\left. 
            +\ b(x)\frac{\partial^2 u(x,t)}{\partial x^2}
           \right)\right]_{x=x^*(t)}
\nonumber\\[8pt]
            &=& b(x^*(t)).
\label{10}
\end{eqnarray}
So indeed, the modal values follow the ODE (\ref{4}).  For an 
ODE with multiple domains of attraction, they
correspond to a multi-modal distribution.  
Furthermore,  
\begin{eqnarray}
    \frac{d}{dt}\left(\frac{\partial^2 u(x^*(t),t)}
            {\partial x^2}\right) &=&
     \left[\frac{\partial^3 u(x,t)}{\partial x^2\partial t}
         +\frac{\partial^3 u(x,t)}{\partial x^3}
         \left(\frac{dx^*(t)}{dt}\right) \right]_{x=x^*(t)}
\nonumber\\[8pt]
    &=&  \left[-\left(\frac{\partial u}{\partial x}\right)\left(
         2\frac{\partial^3 u}{\partial x^3}+\frac{d^2b(x)}{dx^2}
            \right)
         -2\left(\frac{\partial^2u}{\partial x^2}\right)^2
            \right.
\nonumber\\
     && \left. -2\frac{db(x)}{dx}\frac{\partial^2 u}{\partial x^2}
         +\left(\frac{dx^*(t)}{dt}-b(x)\right)
            \frac{\partial^3 u}{\partial x^2}
         \right]_{x=x^*(t)}
\nonumber\\[8pt]
    &=& -2\left(\frac{db(x^*(t))}{dx}+
         \frac{\partial^2 u(x^*(t),t)}{\partial x^2} \right)
    \frac{\partial^2 u(x^*(t),t)}{\partial x^2}.
\label{11}
\end{eqnarray}
Hence, the value $u(x^*(t),t)$ does not change, and
its location follows the ordinary differential
equation $\dot{x}=b(x)$.  A local minimum of the
function $u(x,t)$ follows the corresponding
deterministic ODE.  Furthermore, the curvature about 
it follows Eq. (\ref{11}).

	Since $u(x,t)$ is the rate of convergence of a 
normalized probability distribution for $X^{\epsilon}_t$ 
when $\epsilon\rightarrow 0$, it is a non-negative 
function with its minima necessarily zero.  
In fact, except for the very critical condition 
known as phase transition, the minimum is unique.  
Eqs. (\ref{8}) and (\ref{10}) state that
if an initial $u(x,0)$ is non-negative with a global
minimum zero, $u(x,t)$ will remain non-negative
with global minimum zero.  In other words, the
properties of being a large-deviation rate function 
are preserved.

Finally, observing that the values of local minima
are related to the probability associated with
each ``attractor'', Eq. (\ref{8}) states that diffusion
processes in different ``attractors'' are almost
reducible in the limit of $\epsilon\rightarrow 0$.
Generically speaking, besides the dominant attractor 
with the global minimum of $u(x,t)$, the probability 
of each attractor with a local minimum vanishes. 
This results in the Law of Large Numbers.  However,   
conditioned upon being outside the dominant
attractor, there will be another global attractor.
These states are known as metastable in 
statistical physics.

\section{The Probabilistic Interpretation 
of \boldmath{$u(x,t)$}}

	We now give the precise meaning for $u(x,t)$.
    Let us denote the solution to Eq. (\ref{2}),
with initial condition $f(x,0) = \delta(x-x_o)$.  
As the solution to (\ref{2}), the transition 
probability $f_{\epsilon}(x,t|x_o)$ is
the fundamental solution to the linear PDE.  
For fixed $x$, $t$, and $x_o$,
when $\epsilon\rightarrow 0$, one has
\begin{equation}
     \lim_{\epsilon\rightarrow 0} f_{\epsilon}(x,t)
       = \delta \left(x-x_t\right),
\end{equation}
where $x_t$ is the solution to the ODE
$\dot{x}_t = b(x_t)$ with initial value
$x_o$.   Or in a more authentic probabilistic
notation:
\begin{equation}
    \lim_{\epsilon\rightarrow 0}
    \Pr\left\{ \left| X_t(\epsilon;x_o)-x_t\right|
        > \epsilon \right\} = 0.
\end{equation}
We therefore can introduce
the rate of convergence:
\begin{equation}
    \lim_{\epsilon\rightarrow 0}\
    -\epsilon \ln\frac{\Pr\{x<X_t(\epsilon;x_o)
        \le x+dx
         \}}{dx} = u(x,t;x_o).
\end{equation}

\subsection{Laws of large numbers, central 
limit theorem and theory of large deviations}

Let us consider a sequence of iid (independent, identically
distributed) $X_i$ and their mean value
\begin{equation}
    Z_n = \frac{1}{n}\sum_{i=1}^n X_i.
\end{equation}
The weak law of large numbers (LLN) states that
\begin{equation}
    \lim_{n\rightarrow\infty} Z_n = \mu^*,
        \ \ \ \mu^*=E[X].
\label{eq_16}
\end{equation}
That is, the probability density function for a continuous
random variable
\begin{equation}
    \lim_{n\rightarrow\infty} f_{Z_n}(z)
          = \delta\left(z-\mu^*\right), \ \ \ \mu^*=E[X].
\label{eq_17}
\end{equation}
Furthermore, the central limit theorem states that
\begin{equation}
    \lim_{n\rightarrow\infty}
      \frac{1}{\sqrt{n}}f_{Z_n}
   \left(\frac{x}{\sqrt{n}}+\mu^*\right) =
        \frac{1}{\sqrt{2\pi \sigma^2}}
             e^{-x^2/(2\sigma^2)}, \ \ \ \sigma^2=Var[X].
\label{eq_18}
\end{equation}
We note that the result in Eq. (\ref{eq_18}) implies
Eq. (\ref{eq_17}), which implies Eq. (\ref{eq_16}).

The large-deviation rate function for the LLN given in
Eq. (\ref{eq_16}) is defined as
\begin{equation}
     u(x) = \lim_{n\rightarrow\infty}
              -\frac{1}{n}\ln f_{Z_n}(x).
\end{equation}
Note that the convergence of the probability 
distribution functions in Eq. (\ref{eq_17}) is always
non-uniform.  Asymptotics beyond all orders, therefore,
necessarily enters the theory of large deviations.

According to Chernoff's formula \cite{touchette}, $u(x)$ is
related to the cumulant generating function (CGF),
$\lambda(\theta)$, for the random variable $X$:
\begin{equation}
    \lambda(\theta) = \ln E\left[e^{-\theta X}\right],
\end{equation}
via the Fenchel-Legendre transform
\begin{equation}
        \lambda(\theta) = \sup_{x}\{\theta x -
               u(x)\}.
\end{equation}
Furthermore, we know that the Fenchel-Legendre transform
of $\lambda(\theta)$,
\begin{equation}
         u^*(x) = \sup_{\theta}\{x\theta-
                 \lambda(\theta)\},
\label{eq_22}
\end{equation}
is the affine regularization of $u(x)$.
$u^*(x)$ is a convex function; it is
identical to $u(x)$ in the neighbourhood
of the global minimum of $u(x)$. Cumulant
(or the Thiele semi-invariants) expansion
has recently found applications in 
renormalization-group approach to singular
perturbation \cite{kirkinis}.

From Eq. (\ref{eq_22}) and the basic properties of
the Fenchel-Legendre transform, we have
\begin{eqnarray}
      && \min_{x}u^*(x) = -\lambda(0) = 0,
\\[7pt]
     &&  x\big|_{u^*(x)=0} =
            \lambda'(\theta)\big|_{\theta=0} = E\left[X\right]
               =\mu^*,
\\[7pt]
     &&  \frac{d^2}{dx^2}u^*(x)\Big|_{x=\mu^*} =
         \left(\lambda''(\theta)\big|_{\theta=0} \right)^{-1}
          = \left(Var[X]\right)^{-1} = \frac{1}{\sigma^2}.
\end{eqnarray}
Therefore, we have in the neighbourhood of
global minimum of $u(x)$, $x=\mu^*$:
\begin{equation}
      u(x) = u^*(x) = \frac{(x-\mu^*)^2}{2\sigma^2}
                + a_3 (x-\mu^*)^3 + \cdots.
\label{eq_26}
\end{equation}
While $u^*(x)$ is convex, $u(x)$ need not
be.  Hence, away from the global minimum $x=\mu^*$,
$u(x)$ can have many local minima.  Let
us denote their locations as $x_{\ell}$ and with corresponding
local expansions
\begin{equation}
            u(x) = u(x_{\ell}) +
                      \frac{a_{\ell}}{2}\left(x-x_{\ell}\right)^2
                      +\cdots, \ \ \ \ell=1,2,\cdots
\label{eq_27}
\end{equation}
in which $u(x_{\ell}) > 0$ and $a_{\ell}>0$.
The $f_{Z_n}(x)$ then has an asymptotic expansion
\begin{equation}
     f_{Z_n}(x) =  e^{-nu(x) + o(n)}
                = \sqrt{\frac{n}{2\pi\sigma^2}}\
                   e^{-nu(x) + o(n)}.
\label{eq_28}
\end{equation}

    Eq. (\ref{eq_28}) implies Eq. (\ref{eq_18}) in the
following sense:
\begin{eqnarray*}
        && \int_{-\infty}^{\infty}
        \Big\| f_{Z_n}(x)- \sqrt{\frac{n}{2\pi\sigma^2}}
          e^{-\frac{n(x-\mu^*)^2}{2\sigma^2}} \Big\|\ dx
\\[8pt]
    &=&  \int_{-\infty}^{\infty}
         \Big\| \sqrt{\frac{n}{2\pi\sigma^2}}
          e^{-nu(x)+o(n)}
          - \sqrt{\frac{n}{2\pi\sigma^2}}
          e^{-\frac{n(x-\mu^*)^2}{2\sigma^2}} \Big\|\ dx
\\[8pt]
         &=& \int_{-\infty}^{\infty} \sqrt{\frac{n}{2\pi\sigma^2}}
          e^{-\frac{n(x-\mu^*)^2}{2\sigma^2}}
         \Big\| e^{-na_3((x-\mu^*)^3+o(n)}
            - 1 \Big\|\ dx
\\[8pt]
         &=& \frac{1}{\sqrt{\pi}} \int_{-\infty}^{\infty}
          e^{-z^2}
         \Big\|e^{-\frac{\widetilde{a}_3z^3}
          {\sqrt{n}}+o\left(\frac{1}{\sqrt{n}}\right)}-1\Big\|\ dz
\\[8pt]
          &=& \frac{1}{\sqrt{\pi}} \int_{-\infty}^{\infty}
          e^{-z^2}
         \Big\|  \frac{\widetilde{a}_3z^3}
          {\sqrt{n}}+o\left(\frac{1}{\sqrt{n}}\right) \Big\|\ dz
\\[8pt]
          &=& O\left(\frac{1}{\sqrt{n}}\right).
\end{eqnarray*}
That is, the convergence of the distribution in
Eq. (\ref{eq_18}) is in $L_1$ and it is of the
order $\frac{1}{\sqrt{n}}$.

    Note that the non-convex parts of $u(x)$, as
given in Eq. (\ref{eq_27}), only contribute to terms
on the order of
\begin{equation}
            e^{-nu(x_{\ell})}\int_{-\infty}^{\infty}
             e^{-\frac{na_{\ell}}{2}(x-x_{\ell})^2}\ dx
          = \sqrt{\frac{2\pi}{na_{\ell}}}\ e^{-nu(x_{\ell})}.
\end{equation}
It is exponentially small, i.e., beyond all orders.

\subsection{Linear dynamics, Gaussian processes, and
an exactly solvable HJE}

    We now consider the SDE in (\ref{1}) with linear
drift, i.e., a Gaussian process:
\begin{equation}
    dX_t = -bX_tdt + \sqrt{2\epsilon}dB_t.
\end{equation}
The corresponding Kolmogorov forward equation is
\begin{equation}
    \frac{\partial f(x,t)}{\partial t}
     = \epsilon \frac{\partial^2 f}{\partial x^2}
     + \frac{\partial}{\partial x}\left( bxf(x,t)\right),
\end{equation}
and the WKB ansatz leads to the PDE
\begin{equation}
   \frac{\partial u_{\epsilon}(x,t)}{\partial t}
    = -\left[\left(\frac{\partial u_{\epsilon}}{\partial x}\right)^2
           - bx\frac{\partial u_{\epsilon}}{\partial x} \right]
      + \epsilon\left[\frac{\partial^2 u_{\epsilon}}{\partial x^2}
          - b \right]
\label{32}
\end{equation}
for $u_{\epsilon}(x,t)$.  We can expresses
\begin{equation}
    u_{\epsilon}(x,t) = a(t) + \frac{\left(x-\mu(t)\right)^2}{2\sigma^2(t)},
\label{33}
\end{equation}
then we have a set of nonlinear ODEs:
\begin{subequations}
\label{eq_34}
\begin{eqnarray}
    \frac{d}{dt}\sigma^2(t) &=& 2\left(1-b\sigma^2(t)\right),
\\
    \frac{d\mu(t)}{dt} &=& -b\mu(t),
\\
    \frac{da(t)}{dt} &=&
           +\epsilon\left(\frac{1}{\sigma^2}-b\right).
\end{eqnarray}
\end{subequations}
Their explicit solution is
\begin{subequations}
\begin{eqnarray}
    \sigma^2(t) &=& \frac{1}{b}+\left(\sigma^2(0)-\frac{1}{b}
            \right)e^{-2bt}, 
\\[6pt]
    \mu(t) &=& \mu(0)e^{-bt}, 
\\[6pt]
    a(t) &=& \frac{\epsilon}{2}\ln\left(
        \frac{\sigma^2(t)}{\sigma^2(0)}\right) + a(0).
\end{eqnarray}
\end{subequations}
Assembling these together, we have
\begin{equation}
    e^{-u_{\epsilon}(x,t)/\epsilon} = \frac{A}{\sqrt{2\pi\epsilon\sigma^2(t)}}
    \exp\left\{-\frac{\left(x-\mu(t)\right)^2}{2\epsilon\sigma^2(t)}
            \right\},
\end{equation}
where $A=\sqrt{2\pi\epsilon\sigma^2(0)}e^{-a(0)/\epsilon}$
is a constant.

    We note that in Eq. (\ref{eq_34}), the $\epsilon$ term
only contributes to $a(t)$.  This is a very surprising
result: The diffusive behavior characterized by
$\sigma^2(t)$ actually is determined by a first-order
nonlinear PDE (\ref{7})!  As we
shall see below, the solution to Eq. (\ref{7}) can be
constructed via the method of characteristics.

\subsection{Large-deviation rate function on a circle}

	We now consider dynamics on a circle 
$\theta\in\mathbb{S}^1[0,1]$ with periodic angular
velocity $\dot{\theta}=b(\theta)=b(\theta+1)$.  In general, 
a saddle-node bifurcation on the cycle gives rise 
to a counter-clockwise or clockwise cyclic 
motion when either $b(\theta) > 0$ or $b(\theta) < 0$ 
$\forall \theta\in[0,1]$ \cite{strogatz_book}. 
We have shown above that the 
corresponding large-deviation rate function 
$u(\theta)$ has its minima and maxima corresponding 
to the stable and unstable fixed points.  We now 
illustrate that corresponding to this cyclic motion, 
$u(\theta)$ becomes a constant on the entire circle.

\begin{figure}[h]
\[
\includegraphics[width=4.5in]{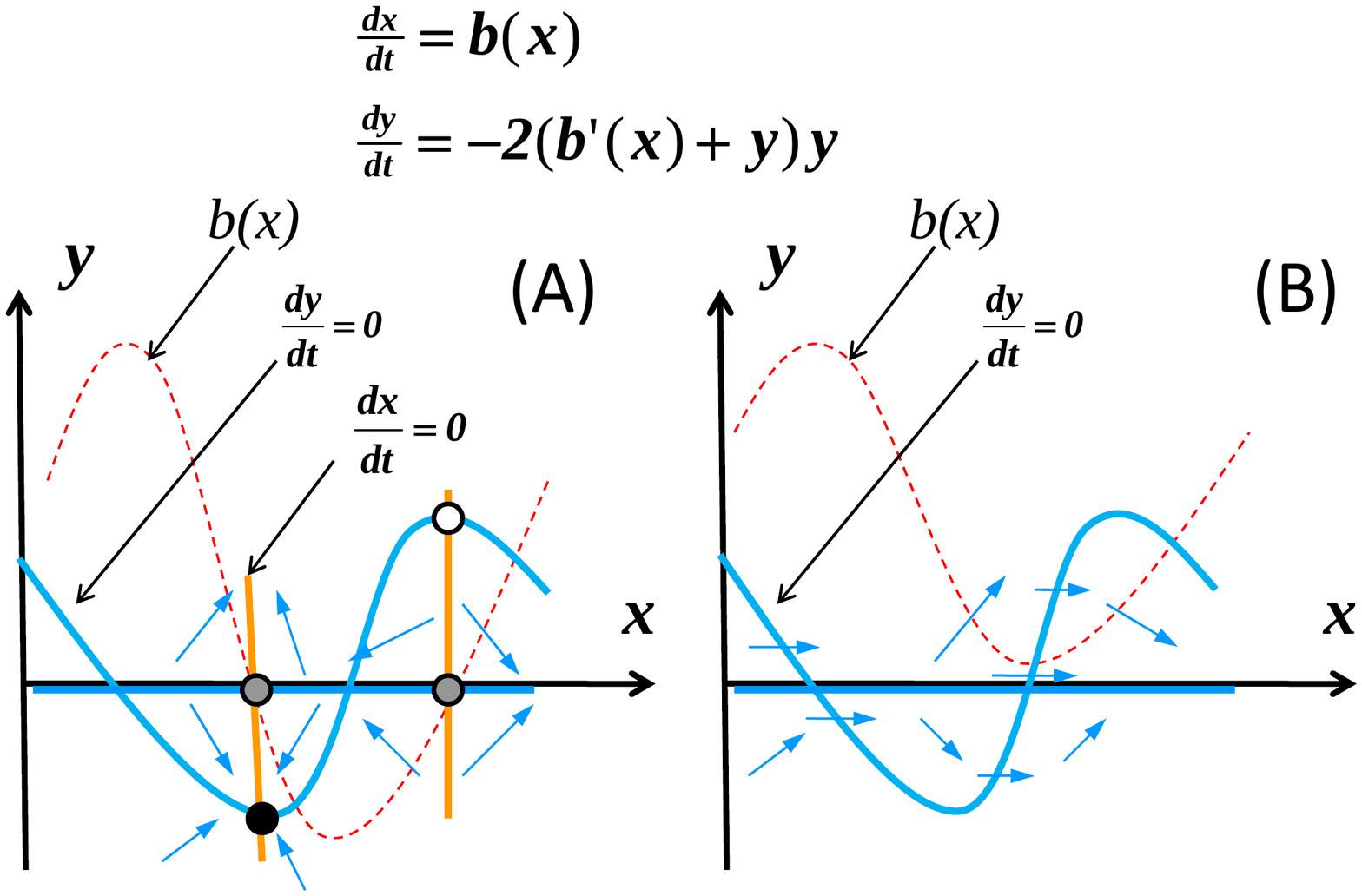}
\]
\caption{The local minimum and its dynamics
of $u(\theta,t)$ on a circle.  $x$ is the location
of a minimum and $y$ its curvature, with their
dynamics given in Eq. (\ref{49}).  When
the dynamical system on the circle has
a stable fixed point at $x^{\dag}$, i.e.,
$b(x^{\dag})=0$ and $b'(x^{\dag})<0$,
the local minimum will approach
$x=x^{\dag}$ and $y=-b''(x^{\dag})$.
The black filled circle in (A) represent a
stable fixed point of the $xy$ planar
dynamics, the grey and open circles represent
saddles and unstable node.  However, when there is
Hopf bifurcation as shown in (B), the
$x(t)$ and $y(t)$ will be periodic functions
of $t$. $u(\theta,t)$ will not reach stationarity.
However, as shown in Eq. (\ref{52}), infinite
time $u(\theta,t)$ is a trival periodic function 
with a constant value. 
}
\label{fig:3}
\end{figure}

	Without losing generality, we shall assume
there is only one minimum and one maximum of $u(x,t)$
and let $x(t)$ and $y(t)$ be the location of
the minimum of $u(\theta,t)$ and the corresponding
curvature.  We are interested in how  $x$ and $y$ 
behave in the infinitely long time limit.
Following Eqs. (\ref{10}) and (\ref{11})
we have
\begin{subequations}
\label{49}
\begin{eqnarray}
    \frac{dx}{dt} &=& b(x),
\\
    \frac{dy}{dt} &=& -2\left(b'(x)+y\right)y.
\end{eqnarray}
\end{subequations}
Fig. \ref{fig:3} graphically shows the occurence of a
Hopf bifurcation in this autonomous planar
system in $\mathbb{R}^2$, corresponding to 
a saddle-node bifurcation of $b(\theta)$ on 
the circle $\mathbb{S}^1$.

We now show for the
case of $\dot{x}=b(x)> 0$ periodic solution
as shown in Fig. \ref{fig:3}B, with $y(t)\rightarrow 0$
asymptotically.  Since $x(t)$ is a periodic function of
$t$, $b'(x(t))$ is also periodic.  Furthermore,
\begin{equation}
    \phi(t) = \int_0^t b'\left(x(s)\right)ds
            = \int_0^t \frac{b'(x(s))}{b(x(s))}
                 \left(\frac{dx(s)}{ds}\right) ds
            =  \ln b(x(t)) + \textrm{const.}
\end{equation}
is also a periodic function of time.  Then
\begin{equation}
     \frac{dy}{dt} = -2\left(\frac{d\phi(t)}{dt}+y\right)y.
\end{equation}
This equation can be re-written as
\begin{equation}
    \frac{d}{dt}\left(y(t)e^{2\phi(t)}\right)
                = -2\left(y(t)e^{2\phi(t)}\right)^2
                e^{-2\phi(t)}.
\label{52}
\end{equation}
The right-hand-side of Eq. (\ref{52}) is $\le 0$ and
it is $0$ iff $y(t)=0$.  Therefore, $y(t)\rightarrow 0$
if $y(0)>0$.  In other words, the curvature 
approaches zero.  The large deviation function
along a limit cycle is a constant \cite{ge_qian_3d}.

\section{A ``Fictitious'' Classical Newtonian Motion}

	Eq. (\ref{7}) has the form of a Hamilton-Jacobi equation
(HJE) for a classical Newtonian motion according to analytical mechanics \cite{goldstein_book,landau_book,arnold_book}. 

\subsection{A Hamiltonian system}

	In analytical mechanics, the solution $u(x,t)$ to
the HJE, a nonlinear PDE like (\ref{7}), is called a 
principal function.  $u(x,t)$ furnishes the entire 
family of orbits, i.e., a flow, corresponding to an 
associated Hamiltonian dynamical system in terms of a 
system of nonlinear ODEs.

	The Hamiltonian system, with time-independent 
Hamiltonian\footnote{The Hamiltonian
associated with a discrete birth-and-death process
(B\&DP) with birth and death rates $u_n$ and $w_n$ is
$H(p,q)=u(q)e^{p}+w(q)e^{-p}-u(q)-w(q)$ where
$u(q)$ and $w(q)$ are the continuous limits of
$u_n$ and $w_n$.  Note this $H(p,q)$ is reduced to
Eq. (\ref{h1}) if $p\ll 1$.  The B\&DP and diffusion 
corresponding to very different forms of the HJE leads to the
diffusion's dilemma \cite{zhou_qian_pre}.
}
\begin{equation}
               H(q,p) = p^2+b(q)p,
\label{h1}
\end{equation}
is a system of autonomous ODEs
\begin{equation}
\label{h1_sys}
    \left\{ \begin{array}{ccl}
            \dot{q} &=& \frac{\partial H}{\partial p}
                    = 2p + b(q),
\\[8pt]
            \dot{p} &=& -\frac{\partial H}{\partial q}
                    = -p \frac{db(q)}{dq}.
\end{array}
 \right.
\end{equation}
It is a re-formulation of a Newton's ``equation of motion''
\begin{equation}
    \frac{d^2q}{dt^2} = b(q)\frac{db(q)}{dq} =
                   \frac{d}{dq}\left(\frac{b^2(q)}{2}\right).
\label{eom}
\end{equation}

This ``fictitious'' analytical mechanical system associated
with stochastic dynamics has been identified and explored 
by Graham and \'{T}el \cite{graham_tel},
and in the even earlier work of Martin-Siggia-Rose \cite{msr_73}.

	According to analytical mechanics, the Hamiltonian system 
also has a corresponding ``Lagrangian''
\begin{equation}
      \mathcal{L}\left[q,\dot{q}\right]
         = \left[p\dot{q}-H(q,p)\right]_{p=\frac{1}{2}
           \left(\dot{q}-b(q)\right)}
         = \left(\frac{\dot{q}-b(q)}{2}\right)^2,
\end{equation}
with the action {\em functional},
\begin{equation}
        S_0\left[q(t);(t_0,q(t_0))\rightarrow(t,q(t))\right]
           =\int_{t_0}^t
           \frac{1}{4}\left[\frac{dq(s)}{dt}-b(q(s))
            \right]^2\ ds.
\label{lagrangian}
\end{equation}
In classical mechanics, the action has never had a 
``meaning'' more than being a mathematical device which 
provides trajectories for a mechanical motion.
In the present work, however, $e^{-S_0/\epsilon}$ turns 
out to be exactly the probability of a path of the
stochastic dynamics in (\ref{1}) with vanishing $\epsilon$.
For finite $\epsilon$, the generalized Onsager-Machlup 
functional is $S_{\epsilon}=S_0+(\epsilon/2)b'(q)$
\cite{onsager,haken,dekker,hunt}.   
The action functional in Eq. (\ref{lagrangian}) 
plays a central role in Freidlin-Wentzell's theory of large
deviations \cite{freidlin_book}.  Also see \cite{zeitouni}
for a rigorous mathematical treatment and \cite{gaspard_book}
for a monograph with an applied mathematical bend. 

One should not confuse the present Hamiltonian system
with a more involved Hamiltonian formulation 
associated with stochastic 
dynamics recently proposed in \cite{xing_jpa_10}.

\subsection{Path integral formalism for the
probability of a diffusion process}
\label{sec:4.2}

	We now show heuristically that the path integral 
in Eq. (\ref{lagrangian}) represents 
the probability of a diffusion process according 
to Eq. (\ref{2}).  Statistical physicists have 
long used it as a useful mathematical tool, 
starting with Onsager and Machlup  
\cite{onsager} for Gaussian processes and
later by Haken, Hunt and Ross \cite{haken,hunt} 
for the general diffusion process.  See \cite{orland} 
for its application in the protein folding problem and
\cite{wang_jcp_10} 
for a very recent study on nonequilibrium
steady-state physics using this formalism.

First, let us consider the solution to a 
simple one-dimensional diffusion process 
defined by Eq. (\ref{1}) with a linear
drift $bx$.  The corresponding
Fokker-Planck equation (\ref{2}) has
the fundamental solution
\[
   f\left(x,t+\Delta t|x',t\right)
 = \frac{1}{\sqrt{4\pi\epsilon\Delta t
\left(\frac{e^{2b\Delta t}-1}{2b\Delta t}\right)}}
    \exp\left[-\frac{\left(x-x'e^{b\Delta t}\right)^2}
          {2(\epsilon/b)\left(e^{2b\Delta t}-1\right)}\right]
\]
\begin{equation}
  \approx \frac{1}{\sqrt{4\pi\epsilon\Delta t
\left(1+b\Delta t\right)}}
    \exp\left[-\frac{(x-x'-bx'\Delta t)^2}
          {4\epsilon\Delta t}\right].
\label{constant_b}
\end{equation}
Now for the solution to the general 
Eq. (\ref{2}) with drift $b(x)$, the
probability density of a trajectory is
approximately
\[
       f\left(x_n,n\Delta t;x_{n-1},(n-1)\Delta t;
       \cdots, x_1,\Delta t|x_0,0\right)
\]

\[   = \frac{1}
      {\sqrt{4\pi\epsilon\Delta t}}\prod_{i=1}^n
    \exp\left[-\frac{(x_i-x_{i-1}-b(x_{i-1})\Delta t)^2}
               {4\epsilon\Delta t}
          -\frac{1}{2}\ln\left(1+b'(x_i)\Delta t\right)\right]
\]

\begin{equation}
  \approx \mathcal{A}\exp\left\{-\frac{1}{4\epsilon}\int_0^t \left[
          \left(\frac{dx(t)}{dt}-b(x(t))\right)^2
              +2\epsilon b'(x(t))
             \right]dt\right\} = \mathcal{A} e^{-\frac{1}{\epsilon}
                  S_{\epsilon}[x(t)]},
\label{path_for}
\end{equation}
where $t=n\Delta t$ and $\mathcal{A}$ is an appropriate
normalization factor.
Note that according to Wiener's theory of 
diffusion, the stochastic trajectory $x(t)$
is nowhere differentiable \cite{gardiner_book}. 
Hence, the use of $dx(t)/dt$ in Eq. (\ref{path_for}) is 
heuristic, and only becomes mathematically 
meaningful in the limit $\epsilon=0$.  
See \cite{zeitouni} and \cite{freidlin_book} for
a rigorous treatment.

\subsection{Conditional probability interpretation
of ``excess kinetic energy''}

A Hamiltonian system has a conserved quantity:
the $H(p(t),q(t))$, along each and every trajectory.  
If we re-arrange the first equation in (\ref{h1_sys}), 
we obtain $p=\left(\dot{q}-b(q)\right)/2$.  Then,
\[
    H(p,q) = p\left(p+b(q)\right)
           = \frac{1}{4}\left(\dot{q}-b(q)\right)
                 \left(\dot{q}+b(q)\right)
           = \frac{1}{4}\left\{\dot{q}^2-
                 b^2(q)\right\},
\]
that is
\begin{equation}
               \dot{q}^2 = b^2(q) + 4H,
\end{equation}
in which $H$ is constant along a trajactory.  
Comparing this with the noiseless trajectory
$\dot{q}=b(q)$ with $H=0$, we see there 
is a constant $4H$ added to the $\dot{q}^2$.
This result can be interpreted as follows.

	We shall call the square of the velocity $\dot{q}^2$ 
the ``kinetic energy''.  A noiseless trajectory, i.e., 
the one with $H=0$, follows the differential equation $\dot{q}=b(q)$.
This means that with starting time $t=t_1$ at position
$q_1$, the noiseless trajectory will be precisely
at $q_2$ at time $t_2$.  Any other 
trajectories arriving at $q_2$ with a different time 
are {\em impossible} for deterministic dynamics, and 
are {\em rare events} when $\epsilon$ is small. 
If, however, one observes such an rare event: 
motion from $q_1$ to $q_2$ with a time $\hat{t}_2\neq t_2$,
what will be the ``most probable trajectory among
all possible $\hat{q}(t)$ with 
$\hat{q}\left(t_1\right)=q_1$ and
$\hat{q}\left(\hat{t}_2\right)=q_2$?

	This is a problem of conditional probability.
Among all the rare trajectories,
\[     \{ \hat{q}(t)|\hat{q}(t_1)=q_1, 
            \hat{q}\left(\hat{t}_2\right)=q_2,\hat{t}_2\neq t_2
    \},
\]
the most probable one follows a solution to the 
Hamiltonian system with an appropriate $H\neq 0$!

More explicitly, since $\hat{q}(t)$ arrives at $q_2$ with 
a time different from $t_2$, say $\hat{t}_2<t_2$, then 
it has to be speeded up compared to $q(t)$.  On the other hand, if $\hat{t}_2>t_2$, it has to be slowed down.  The solution 
to the Hamiltonian equation states that the most 
probable trajectory is the one having a {\em constant} 
amount of {\em excess kinetic energy} ($4H$) added or 
reduced along the trajectory.  In the theory of probability,
this is a consequence of the van Campenhout-Cover
theorem \cite{ge_qian_info}.  

	The fictitious Hamiltonian system, therefore, is the 
consequence of stochastic dynamics conditioned on the 
occurence of a rare event.  The kinetic energy exists 
``retrospectively'' in some stochastic dynamics.

\section{Solutions to the Hamilton-Jacobi Equation (HJE)}

Since the HJE is a nonlinear PDE, there is no
systematic method to obtain its solution for
arbitary initial data.  Rather, there are
classes of solutions one can obtain;
exactly or approximately.  Since this is a rather
developed area of applied mathematics, we 
shall only touch upon some issues highly
relevant to our mission. 

\subsection{A class of exact solutions}

    The standard way to solve a HJE is precisely
by solving its corresponding Hamiltonian dynamics in
terms of the ODEs.  Since the Hamiltonian $H(q,p)$
does not depend explicitly on time, one can
verify that $u(x,t)=u_0(x)-Et$ is a
solution to HJE (\ref{7}) \underline{if} $u_0(x)$ is
a solution to
\begin{equation}
    H\left(\frac{du_0(x)}{dx}, x\right)
              = E.
\label{eq_46}
\end{equation}
Therefore,
\begin{equation}
    u(x,t) = \int_{x_0}^x p(q)dq - Et
    = \frac{1}{2}\int_{x_0}^x
            \left(-b(z)\pm\sqrt{b^2(z)+4E}\right)dz
            -Et.
\label{eq_47}
\end{equation}
Initial data $u_0(x)$ satisfying Eq. (\ref{eq_46})
is called {\em characteristic initial data} \cite{evans_book}.

    For $E=0$, $u(x,t)=u_0(x)$ $\forall t$.
Hence it is a stationary solution to the HJE.  In
this case, Eq. (\ref{eq_46}) yields
\begin{equation}
           u^{st}(x)=-\int_0^x b(z)dz
\label{eq_48}
\end{equation}
and a trivial solution $u_0(x)$.
The solution in Eq (\ref{eq_48}) corresponding
to $E=0$ is the expected stationary solution.

\subsection{Solutions via characteristics}

    The solutions given in Eq. (\ref{eq_47})
is in a special class.  Certainly not any initial
data $u(x,0)$ is {\em characteristic}.
More importantly, the $u(x,t)$ in (\ref{eq_47})
can not satisfy the basic properties of a large
deviation rate function: It has to be
non-negative with its minimum exactly
being zero.  Therefore, we need to look for
other possible solutions to the HJE (\ref{7})
corresponding to noncharacteristic initial data
\cite{evans_book}.

According to Evans \cite{evans_book},
let
\[    \frac{\partial u}{\partial t} \mapsto y, \ \
      \frac{\partial u}{\partial x} \mapsto z, \ \
      u \mapsto u, \ \
      t \mapsto t, \ \ x \mapsto x.
\]
The the HJE (\ref{7}) is in the form of
\[
          F(y,z,u,t,x)=y+z^2+zb(x) = 0,
\]
with
\begin{subequations}
\begin{eqnarray}
    \dot{y}(s) &=& -F'_t = 0,
\\
    \dot{z}(s) &=& -F'_x = -z\frac{db(x)}{dx},
\\
    \dot{u}(s) &=& F'_yy + F'_zz = y+\left(2z+b(x)\right)z,
\\
    \dot{t}(s) &=&  F'_y(y,z,t,x) = 1,
\\
    \dot{x}(s) &=& F'_z(y,z,t,x) = 2z+b(x).
\end{eqnarray}
\label{from_evans}
\end{subequations}
Note that $y=-(z^2+zb(x))$ is the Hamiltonian $-H(x,z)$.
The pair of equations (\ref{from_evans}b) and
(\ref{from_evans}e) are the Hamiltonian system in Eq. (\ref{h1_sys}). $\dot{u}(s)=\dot{x}z-H(x,z)$ is the
corresponding ``Lagrangian''.

\begin{figure}[t]
\[
\includegraphics[width=5in]{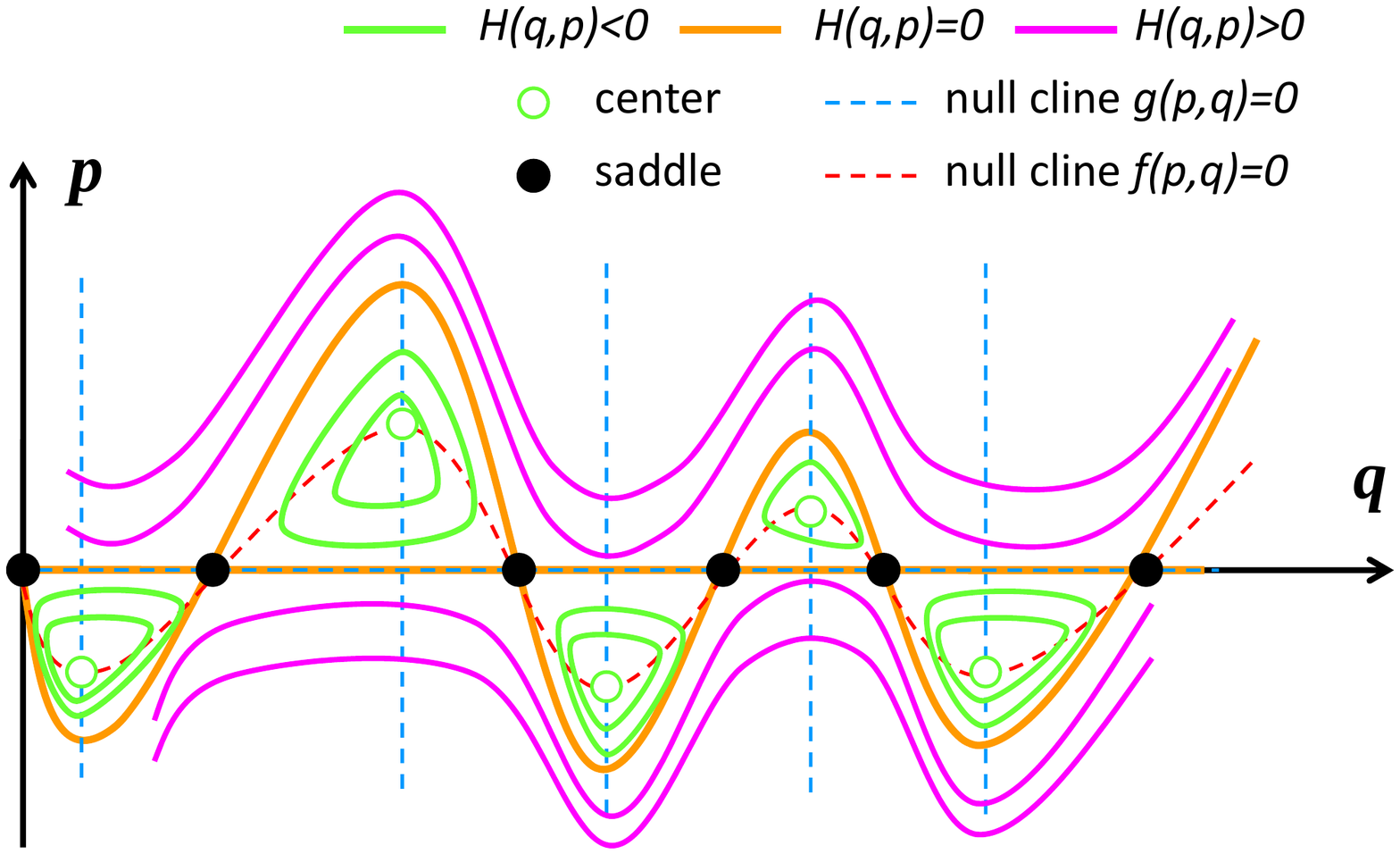}
\]
\caption{Phase portrait for Hamiltonian
system (\ref{h1_sys}): $\dot{q}=f(p,q),
\dot{p}=g(p,q)$.  The nullcline for
$g(p,q)=0$ are $p=0$ and $q=q^*$
where $q^*$s are where $db(q)/dq=0$,
shown in blue dashed lines.  The nullcline
for $f(p,q)=0$ are $p=-b(q)/2$, shown
in red dashed lines.  The orange lines
of $p=0$ and $p=-b(q)$ are the contour
for $H(q,p)=p^2+b(q)p=0$.  The intersections
between blue and red dashed lines are equilibrium
points of the Hamiltonian system: Centers are
by green open circles and saddles are in black
filled circles.  Trajectories in green have $H<0$
and in pink have $H>0$.
}
\label{fig:2}
\end{figure}

\subsection{Phase portrait of characteristic
lines of Hamiltonian system}

In our case, Eq. (\ref{h1_sys}) can be solved since
$p^2+b(q)p=E$ is a constant of motion:
\begin{equation}
    \frac{dq}{dt} = \pm\sqrt{b^2(q)+4E},  \ \
          \textrm{i.e.,} \
         \int \frac{dq}{\sqrt{b^2(q)+4E}}
            = \pm t + \textrm{const.}
\label{eq_480}
\end{equation}
When $E=0$, this is precisely the solution to the
ODE $dq/dt=\pm b(q)$.  Fig. \ref{fig:2} shows
the phase portrait of the Hamiltonian dynamics.

    The phase portrait in term of $p$ as a
function of $q$ is
\begin{equation}
    p(q) = \frac{-b(q)\pm\sqrt{b^2(q)+4E}}{2}.
\end{equation}

\subsection{Solution to HJE with noncharacteristic
initial value}

 As a large-deviation rate function, one
appropriate initial condition for $u(x,t)$ should
be $u(x,0)=0$.  It is clear
that $u(x,t)=0$ is a solution to the HJE, but
this is not a meaningful one.  Therefore,
one could be interested in the solution to the
HJE with an infinitesimal initial data $u(x,0)$.
Therefore, initially, one can linearize Eq.
(\ref{6}):
\begin{equation}
       \frac{\partial u_0(x,t)}{\partial t}
    = -b(x)\frac{\partial u_0}{\partial x}
      + \epsilon\left[\frac{\partial^2 u_0}{\partial x^2}
          + \frac{db(x)}{dx}\right] + O\left(\epsilon^2\right).
\end{equation}

    The exact result for a Gaussian process can provide
some insights:  Note that if $\sigma^2(0)=\infty$,
then the initial $u(x,0)=0$.

\subsection{HJE on a circle}

We again consider $\theta\in\mathbb{S}^1[0,1]$ 
with periodic angular velocity 
$\dot{\theta}=b(\theta)$.  The $b(\theta)$ on
a circle can be decomposed as
\begin{equation}  
            b(\theta) = b_0 
             -\frac{dU(\theta)}{d\theta}, 
\end{equation}
with a differentiable periodic potential 
$U(\theta)=U(\theta+1)$, $\theta\in\mathbb{S}^1$.
The corresponding Hamiltonian equation on a torus is
\begin{equation}
    \left\{ \begin{array}{ccl}
         \dot{\theta} &=& \displaystyle   
           2\omega + b(\theta) 
               \ = \ 
                2\omega + b_0 
                - \frac{dU(\theta)}{d\theta},
\\[8pt]
            \dot{\omega} &=& \displaystyle
            -\omega \frac{db(\theta)}{d\theta}
              \ = \ \omega\frac{d^2U(\theta)}{d\theta^2},
\end{array}
 \right.
\label{fhs_on_S}
\end{equation}
with the Hamiltonian 
$H(\theta,\omega)=\omega^2+\omega b_0-\omega U'(\theta)$;
$(\theta,\omega)\in\mathbb{S}^2$.

	In general, there are two types of fixed points
$(\theta,\omega)^*$ in system (\ref{fhs_on_S}):
$\omega^*=0$ and $U'(\theta^*)=b_0$, 
or $U''(\theta^*)=0$ and 
$\omega^*=(1/2)(U'(\theta^*)-b_0)$.
These two types are the same for systems with $b_0=0$
(i.e., a gradient system on $\mathbb{S}^1$).  For
a rotational system with $b(\theta)>0$ or $<0$ on the
entire $\mathbb{S}^1$, the first type does not exist. 
Therefore, taking the index theory for a $C^1$-vector in
a plane into consideration \cite{perko_book}, the 
occurence of a cyclic motion in $\dot{\theta}=b(\theta)$ 
with changing $b_0$ corresponds to an annihilation of two 
fixed points of the two types.

%The linear analysis near a fixed point yields corresponding
%matrices for the two types, respectively: 
%	\left( \begin{array}{cc}
%        -U''(\theta^*) & 2  \\[7pt]  0 & U''(\theta^*)
%              \end{array} \right), \ \
% 	\left( \begin{array}{cc}
%        0 & 2  \\[7pt]  \frac{1}{2}
%                \left(U'(\theta^*)-b_0\right)U'''(\theta^*) 
%                        & 0
%              \end{array} \right).
%\end{equation}
%Since there could be no node nor spiral in a Hamiltonian 
%system, the first type has to be a saddle point and the 
%second type is a center when 
%$\left(U'(\theta^*)-b_0\right)U'''(\theta^*)<0$ and
%a saddle when 
%$\left(U'(\theta^*)-b_0\right)U'''(\theta^*)>0$.

\section{High Dimensional Cases: Momentum,
Entropy Production and Nonequilibrium}

	From a standpoint of the theory of Markov
processes, the one-dimensional system in (\ref{1}) 
can only reach a time-reversible stationary process
\cite{jqq_book}.  This result corresponds to
the statement that an ODE on $\mathbb{R}^1$ is
always a gradient system: $\dot{x} = b(x) = -dU(x)/dx$
with $U(x)=-\int b(x)dx$.  For autonomous ODE systems 
in higher dimensions, a limit cycle can occur.  This
corresponds to nonequilibrium phenomena in 
system (\ref{eq_55}) in dimension 2 or higher.

	We now consider an $N$-dimensional 
diffusion process with the stochastic differential
eqaution
\begin{equation}
     dX_t = b(X_t)dt + \sqrt{2\epsilon} dB_t,
 \ \ \ X_t\in\mathbb{R}^N,
\label{sde_nd}
\end{equation}
and a corresponding Fokker-Planck equation for 
the probability density function $f(x,t)dx$
$=$ $\Pr\{x\le X_t<x+dx\}$,
\begin{equation}
	\frac{\partial}{\partial t} f(x,t)
        = \epsilon\frac{\partial^2}{\partial x^2} f(x,t)
         -\frac{\partial}{\partial x}\left(
             b(x) f(x,t)\right).
\label{fpe_nd}
\end{equation}

\subsection{HJE in two-dimensional systems}

For a two-dimensional SDE:
\begin{equation}
		dX_t = b_x(X_t,Y_t)dt+\sqrt{2\epsilon} dB_t^{(1)}, \ \ 
       dY_t = b_y(X_t,Y_t)dt+\sqrt{2\epsilon} dB_t^{(2)},
\label{eq_55}
\end{equation}
the corresponding fictitious Hamiltonian is
\begin{equation}
      H(q_x,q_y,p_x,p_y) = p_x^2+p_y^2+b_x(q_x,q_y)p_x
                         +b_y(q_x,q_y)p_y,
\end{equation}
so the Hamiltonian dynamical system is
\begin{equation}
	\left\{\begin{array}{ccccl}
	\displaystyle \frac{d}{dt}q_x &=& 
    \displaystyle \frac{\partial H}{\partial p_x}
             &=& 2p_x+b_x(q_x,q_y),
\\[17pt]
   \displaystyle \frac{d}{dt} q_y &=& 
   \displaystyle \frac{\partial H}{\partial p_y} 
             &=&  2p_y+b_y(q_x,q_y),
\\[17pt]
	\displaystyle \frac{d}{dt}p_x &=& 
    \displaystyle -\frac{\partial H}{\partial q_x} 
            &=& 
    \displaystyle -p_x\frac{\partial b_x}{\partial q_x}
                 -p_y\frac{\partial b_y}{\partial q_x},
\\[17pt]
	\displaystyle\frac{d}{dt}p_y &=& 
    \displaystyle -\frac{\partial H}{\partial q_y} 
             &=&  
    \displaystyle -p_x\frac{\partial b_x}{\partial q_y}
                -p_y\frac{\partial b_y}{\partial q_y}.
\end{array} \right.
\end{equation}
Its corresponding Lagrangian is
\begin{equation}
  \mathcal{L}\left[q_x,q_y,\dot{q}_x,\dot{q}_y\right]
           =\left(\frac{\dot{q}_x-b_x}{2}\right)^2+
             \left(\frac{\dot{q}_y-b_y}{2}\right)^2,
\end{equation}
and the equations of motion are
\begin{eqnarray}
       \ddot{q}_x &=& \frac{\partial}{\partial q_x}
                 \left(\frac{b_x^2+b_y^2}{2}\right)
                 - \left(\frac{\partial b_y}{\partial q_x}
			        - \frac{\partial b_x}{\partial q_y}\right)
                          \dot{q}_y
\nonumber\\[-5pt]
\label{eq_58}\\[-5pt]
	  \ddot{q}_y &=& \frac{\partial}{\partial q_y}
                 \left(\frac{b_x^2+b_y^2}{2}\right)
                 + \left(\frac{\partial b_y}{\partial q_x}
                 -\frac{\partial b_x}{\partial q_y}                     
                  \right)
                          \dot{q}_x.
\nonumber 
\end{eqnarray}
In vector form Eq. (\ref{eq_58}) can be written as
\begin{equation}
   \frac{d^2}{dt^2}\vec{q}(t) = \left(\nabla\times\vec{b}
             \right)\times \frac{d}{dt}\vec{q}(t)
            +\frac{1}{2}\nabla \|\vec{b}\|^2. 
\label{eq_59}
\end{equation}
A Lorentz magnetic force like term arises if the vector field
$\vec{b}(q_x,q_y)$ is non-conservative.  

For a system of arbitary dimension, the vector form 
Eq. (\ref{eq_59}) is still valid if one interprets 
\[
        \left(\nabla\times\vec{b}\right)\times
                   \dot{\vec{q}}(t) \ \rightarrow\
		\sum_j\left(\frac{\partial b_i}{\partial q_j}
                   -\frac{\partial b_j}{\partial q_i}
			\right)\frac{dq_j(t)}{dt}.
\]

\subsection{Force decomposition, momentum
and entropy production}

	We now show an interesting relation between the
non-gradient $b(x)$ and the large deviation 
rate function $u^{st}(x)$ for the stationary diffusion.

We consider the case of only a single attractive domain 
with only one stable fixed point at $0$. Suppose the
$N$-dimensional vector field $b(x)$ admits an 
orthogonal decomposition \cite{graham_tel}:
\begin{equation}
    b(x)=-\nabla U(x)+\ell(x),
\end{equation}
where the function $U(x)$ is
continuously differentiable, $\nabla U(x)\neq
0$ for $x\neq 0$, and the inner product in $\mathbb{R}^N$,
$\ell(x)\cdot\nabla U(x)=0$. Then the large
deviation rate function $u^{st}(x)=U(x)+$const., 
and the unique extreme of
the action functional $S_0[x(t): (t_0,0)\rightarrow (T,x^*)]$ is given
by the equation
$$\frac{dx}{dt}=\nabla U(x)+\ell(x),$$
where $t\in [t_0,T]$ and $x(t_0)=0$, $x(T)=x^*$.

Recall that $u^{st}(x)$ is the stationary solution to the
Hamiltonian-Jacobi equation
\begin{equation}
      \frac{\partial u(x,t)}{\partial t}=-\|\nabla u(x,t)\|^2-b(x)\cdot \nabla u(x,t),
\label{hje_hd}
\end{equation}
associated with Hamiltonian $H(q,p)=\|p\|^2+b(q)\cdot p$.

Hence the Hamiltonian dynamics follow
\begin{equation}
\label{fhs_nd}
    \left\{ \begin{array}{ccl}
            \dot{q_i} &=& \displaystyle 
              \frac{\partial H}{\partial p_i}
                    \ = \ 2p_i + b_i(q)
\\[8pt]
            \dot{p_i} &=& \displaystyle
              -\frac{\partial H}{\partial q_i}
                    \ = \ -p\cdot \frac{\partial b(q)}{\partial q_i}
\end{array}
 \right.
\end{equation}
The action functional $S_0$ is just the path integration of the
corresponding Lagrangian
$$\mathcal{L}[\dot{q},q]=[p\dot{q}-H(q,p)]=\sum_i \left (
\frac{\dot{q_i}-b_i(q)}{2}\right )^2.$$

It is easy to calculate that along the classical trajectory associated with $dx/dt=\nabla U(x)+\ell(x)$, we have
\begin{eqnarray}
     p(t)&=&-\nabla u^{st}(x(t));
\nonumber\\[-5pt]
\label{eq_64}
\\[-5pt]
     \dot{q}(t)&=&\nabla U(q(t))+\ell(q(t)).
\nonumber
\end{eqnarray}
The Hamiltonian of this optimal dynamics is always zero, and the
associated trajectories cross all the fixed points of the
deterministic dynamic system (\ref{4}).

	In applied stochastic dynamics, the stationary solution
$u^{st}(x)$ to Eq. (\ref{hje_hd}) can be considered as
a ``landscape'' for the 
dynamics \cite{ge_qian_jrsi,zhou_qian_pre}.
Then Eq. (\ref{eq_64}) provides a very novel ``meaning'' 
for the conjugate momentum in the fictitious Hamiltonian system: 
It is the force associated with the landscape.
Moreover, $(p(t),\ell(q(t))=0$, i.e., the Lorentz
force $\ell(x)$ is perpendicular to the momentum.

For a multi-dimensional diffusion process (\ref{sde_nd})
the entropy production rate for the stationary
diffusion process is defined as \cite{jqq_book}:
\begin{equation}
        e_p=\frac{1}{\epsilon}
 \int\|b(x)-\epsilon\nabla\log\pi(x;\epsilon)\|^2
         \pi(x;\epsilon)dx,
\label{ep}
\end{equation}
in which $\pi(x;\epsilon)$ is the probability 
density function for the stationary process. 
Therefore, $e_p=0$ if and only if $\ell(x)=0$.
$e_p\neq 0$ is widely considered to be 
a fundamental property of a nonequilibrium
steady state; $e_p=0$ implies that the stationary
process $X_t$ is time reversible \cite{jqq_book}.

When $\epsilon$ tends to zero, if $\epsilon\nabla\log
\pi(x;\epsilon)$ converges, then the $u^{st}(x)$ is
related to the stationary probability density 
through a Boltzmann-like relation 
\begin{equation}
          \lim_{\epsilon\rightarrow 0}
      \epsilon\nabla\log\pi(x;\epsilon)=-\nabla u^{st}(x),
\end{equation}
and then
\begin{equation}
   \|b(x)-\epsilon\nabla\log\pi(x;\epsilon)\|^2
                 \rightarrow \|\ell(x)\|^2,
\end{equation}
the left-hand-side of which is inside the integral 
in Eq. (\ref{ep}).  Therefore, asymptotically we
have
\begin{equation}
   e_p \approx \frac{1}{\epsilon}\int
           \|\ell(x)\|^2e^{-U(x)/\epsilon}dx.
\end{equation}

	Cases with multiple attractive domains are
much more complex due to the problem of turning
points and boundary layers in the limit of $\epsilon$
tending to zero \cite{omalley,maier_stein}.

\subsection{Processes with time-reversal}

We now consider the diffusion process corresponding 
to Eqs. (\ref{sde_nd}) and (\ref{fpe_nd}) with a 
time-reversal.  The corresponding forward equation for the 
reversed process is
\begin{equation}
	\frac{\partial}{\partial t} \tf(x,t)
        = \epsilon\frac{\partial^2}{\partial x^2} \tf(x,t)
         -\frac{\partial}{\partial x}\left[\left(
		2\epsilon\frac{\partial\ln\pi (x)}{\partial x}-
		b(x)\right)\tf(x,t)\right],
\end{equation}
where $\pi(x)$ is the stationary density for Eq.
(\ref{fpe_nd}).  In other words, the corresponding drift
is
\begin{equation}
         \tb(x) = 2\epsilon\frac{\partial\ln\pi (x)}
              {\partial x}-b(x).
\end{equation}
In fact, both $b(x)$ and $\tb(x)$ can be
written as
\begin{equation}
     b(x) = \epsilon\frac{\partial\ln\pi(x)}{\partial x}
             +\ell(x), \ \ \
     \tb(x) = \epsilon\frac{\partial\ln\pi(x)}{\partial x}
             -\ell(x), 
\end{equation}
with
\begin{equation}
        \ell(x) = b(x)-\epsilon\frac{\partial\ln\pi(x)}{\partial x}.
\end{equation}
If we denote the stationary distribution
\begin{equation}
    \pi(x) = e^{-U(x)/\epsilon},
\end{equation}
then $b(x)=-\nabla U(x)+\ell(x)$ and 
$\tb(x)=-\nabla U(x)-\ell(x)$.  Their 
corresponding fictitious Hamiltonians are
\begin{equation}
          H(p,q) = p^2+p\left(-\nabla U(q)+\ell(x)\right)
\end{equation}
and
\begin{equation}
       \widetilde{H}(p,q) = p^2+p\left(-\nabla U(q)
           -\ell(q)\right).
\end{equation}
The corresponding ``equations of motion'' are given
by  Eq. (\ref{eq_59}):
\begin{equation}
   \frac{d^2}{dt^2}\vec{q}(t) = \pm\left(\nabla\times
             \vec{\ell}(\vec{q})
             \right)\times \frac{d}{dt}\vec{q}(t)
            +\frac{1}{2}\nabla\left( \|\nabla U(\vec{q})\|^2
                  +\|\vec{\ell}(\vec{q})\|^2\right). 
\end{equation}
In the case of reversible processes, $\ell(x)=0$ and 
$\tb(x)=b(x)=-\nabla U(x)$.

\section{Interpretive Remarks}

	Stochastic dynamics following trajectories 
defined by Eq. (\ref{sde_nd}), with the time-evolution
of the corresponding probability distribution 
characterized by Eq. (\ref{fpe_nd}), has a continuous 
but nowhere differentiable trajectory.  Diffusion
represents motions of a particle in a highly 
viscous medium in which inertia is completely lost 
instantaneously \cite{wax_book}.  At any given position
$x\in\mathbb{R}^N$, the particle can move in any possible 
direction with any possible speed, but the mean velocity 
is $\langle dx\rangle/dt = b(x)$.  
There is no momentum in the classical sense. 

	The analytical mechanical structure 
hidden in the stochastic dynamics discussed in
the present review, however, is Newtonian.
We note that the acceleration for deterministic
dynamics
\begin{equation}
            \frac{dx}{dt} = b(x),
\label{eq_82} 
\end{equation}
is
\[
    \frac{d^2x_i}{dt^2} = \frac{d}{dt}b_i\left(x(t)\right)
              = \sum_{j=1}^N \frac{\partial b_i(x)}{\partial x_j}
                      b_j(x) 
\]
\[
    = \sum_{j=1}^N b_j(x)\left(
         \frac{\partial b_i(x)}{\partial x_j}   
       -\frac{\partial b_j(x)}{\partial x_i}\right) 
         + \sum_{j=1}^N b_j(x)\frac{\partial b_j(x)}{\partial x_i}
\]
\begin{equation}
   =  \left(\nabla\times b(x)\right)\times b(x)
           + \frac{1}{2}\nabla\|b(x)\|^2.
\label{eq_83}
\end{equation}
This is exactly Eq. (\ref{eq_59}).  But how should one
interpret the inertia and momentum in the fictitious 
Newtonian motion?  How can we interepret the Hamiltonian as 
a conserved quantity in a stochastic trajectory?

{\bf\em Mean dynamics and probability moment closure 
problem.}
The mean behavior of nonlinear stochastic dynamics
like (\ref{sde_nd}), while being a deterministic 
function of time, can not be represented by a simple
ordinary differential equation.  For a stochastic
process $x(t)$:
\[    \frac{d}{dt}\langle x(t)\rangle =
           \langle b(x)\rangle \neq
           b\left(\langle x(t)\rangle\right). 
\]
A traditional approach to resolve this problem is to 
introduce higher-order moments for the distribution of $x(t)$, 
and to express the stochastic dynamics in terms of the mean, variance, third moments, etc.  The method of moment closure 
was introduced to reduce this infinite hierarchy to
approximately a finite system.

	Such an approach encounters significant difficulties
if a nonlinear $b(x)$ has multiple domains of attraction.
In this case, the mean dynamics reflects two fundamentally
different behaviors on very different time scales:
the intra-attractor dynamics and inter-attractor dynamics
\cite{qian_nonl}.  For a small $\epsilon$, the latter are
rare events.

{\bf\em Kinetic energy without momentum.}
The equations in (\ref{eq_82}) and (\ref{eq_83})
seem to suggest another line of hierachical characterization
of stochastic dynamics.  The Hamiltonian
dynamics corresponding to Eq. (\ref{eq_83}) 
yields Eq. (\ref{eq_82}) when $H=0$.  For
all other dynamics with $H\neq 0$, they
are impossible for Eq. (\ref{eq_82}), i.e.,
Eq. (\ref{sde_nd}) with $\epsilon=0$.  However,
they are the dynamics of a rare event when
$\epsilon\neq 0$; because they are the
most probable trajectory conditioned upon 
the rare event being observed.  In fact,
the $H$ is the uniform amount of excess kinetic
energy required, added or reduced, to lead
a rare event to occur.  The Hamitonian
dynamics, in this sense, has no forward predictive 
power based on given position and momentum; 
but it can predict detailed
dynamical paths retrospectively, as in the
Lagrangian formulation of classical mechanics
and Fermat's principle for optics.

It is easy to see from Eq. (\ref{eq_83}) that one 
conserved quantity in dynamics is $\dot{q}^2-\|b(q)\|^2$.
Hence, such a solution has a ``uniform excess kinetic 
energy'' compared with Eq. (\ref{eq_82}). Conditioned
on a given rare event, the problem of most probable 
ensemble is precisely the subject of Boltzmann-Gibbs'
statistical mechanics and the Gibbs conditioning in 
the theory of large 
deviations \cite{feng_kurtz,ge_qian_info}.

{\bf\em Evolution of the landscape.}
The large-deviation rate function
$u^{st}(x)$, as the stationary solution to 
the HJE (\ref{hje_hd}) or the ``Boltzmann
factor'' for the stationary solution to
Eq. (\ref{fpe_nd}):
\[   u^{st}(x) = \lim_{\epsilon\rightarrow 0}
                 -\frac{1}{\epsilon}\ln f^{st}(x;\epsilon),
\]
has been known for
a long time to have a Lyapunov
property for the dynamics $dx/dt=b(x)$:
\[
     \frac{d}{dt} u^{st}(x(t)) = \nabla u^{st}(x)
        \cdot b(x) = -\|\nabla u^{st}(x)\|^2 \le 0.
\]
This means the ergodic, stationary stochastic 
dynamics contains a great deal of information on the 
time-dependent behavior of the system.  In
physics and biology, there is a growing 
interest to use the function $u^{st}(x)$ as
an analytical visualization tool for global 
behavior of complex dynamics: a landscape
\cite{kubo,graham_tel,yin_ao,ge_qian_jrsi,zhou_qian_pre}.  
$u^{st}(x)$ can even exist for vector fields $b(x)$ 
which are non-conservative.  The time-dependent HJE
(\ref{hje_hd}), therefore, can be interpreted
as the evolution of the landscape.

\section{Acknowledgement}

We thank Ping Ao, Bernard Deconinck, Jin Feng, 
Robert O'Malley, Vipul Periwal, David Saakian and Jin Wang 
for helpful discussions.

\end{document}